\documentclass[12pt]{article}

\usepackage[margin=1in]{geometry}
\usepackage{url}
\usepackage{graphicx}
\usepackage{hyperref}
\usepackage{amsmath}

\newcommand{\IR}{\ensuremath{\mathit{IR}}}
\newcommand{\CS}{\ensuremath{\mathit{CS}}}
\newcommand{\DP}{\ensuremath{\mathit{DP}}}

\title{Maximizing the Expected Value of a Lottery Ticket: How to Sell and When to Buy}
\author{Allen Kim and Steven Skiena \\
	Dept. of Computer Science \\
	Stony Brook University \\
	\{allekim$\mid$skiena\}@cs.stonybrook.edu}
\date{January 2, 2020}

\begin{document}
\maketitle

\abstract{
Unusually large prize pools in lotteries like Mega Millions and Powerball attract
additional bettors, which increases the likelihood that multiple winners will have to share the pool.
Thus, the expected value of a lottery ticket decreases as the probability of collisions (two or more bettors with identical winning tickets) increase.
We propose a way to increase the expected value of lottery tickets by minimizing collisions, while preserving the independent generation necessary in a distributed point-of-sales environment.
Our approach involves partitioning the ticket space among different vendors and pairing them off to ensure no collisions among pairs.
Our analysis demonstrates that this approach increases the expected value each ticket, without increasing the size of the prize pool. We also analyze when ticket sales have maximal expected value, and show that they provide positive returns when the jackpot is between \$775.2 million and \$1.67 billion dollars.
}

\section{Introduction}\label{sec:intro}

With players attracted by the potential winnings from enormous lottery pools,
multistate lotteries like Mega Millions and Powerball sell tens to hundreds of
millions of tickets each week across the United States.
Larger lottery pools attract more sales, but the real expected value of a particular lottery ticket is a function of combinatorics, pool size, and consumer behavior.

Calculating the probability of winning a lottery is a standard exercise in
combinatorics \cite{on_the_lottery_problem, barboianu_2009, ed1, ed2}.
Each ticket for the Powerball lottery
contains six numbers, with five ``white balls'' in range from 1 to 69, plus one ``powerball'' in the range from 1 to 26.
The grand prize requires selecting all of these numbers correctly.
Thus there are
\[ \binom{69}{5} \binom{26}{1} = 292,201,338 \]
possible tickets, making for very low odds of winning the jackpot.
Similar rules hold for Mega Millions and other lottery games. 

The payoff for a winning ticket varies each week, depending on the size of the prize pool.
The pool for Powerball starts at \$40 million, but increases
each week until there is a winner.
On January 13, 2016 it reached a record high of \$1.586 billion.

Players like to win, but they do not like to share.
Even the largest lottery pool will yield a disappointing payoff if too many players independently select the winning ticket.
And multiple winners become a genuine risk once the pools get big enough \cite{megamill}.
Indeed, fully six of the ten biggest Powerball/MegaMillions jackpots (as of
March 28, 2019) had multiple winners.
Two of these big jackpots had to be shared three ways.

In this paper, we propose and analyze a practical scheme to increase the
likelihood of single winners, or equivalently to minimize the probability of sharing.
Paradoxically, this manages to increase the expected value of a lottery
ticket {\em without} costing the central authorities any additional contributions to the
payoff pool.
Given that larger potential winnings attract more players, we anticipate that
implementation of our scheme would generate increased interest in these games,
and enlarge the ostensible benefits the governments running them can provide.

Further, we demonstrate that the number of Powerball tickets bought increases
quadratically with pool size, which implies that tickets become increasing {\em less} valuable after the pool passes a critical threshold.
This analysis enables us to determine the range of pool sizes where tickets have positive expected value.
In particular, we establish that Powerball tickets bought (under the current sales model) with
pool sizes between \$775.2 million and \$1.6656 billion have positive expected value.

\section{Selecting Lottery Tickets with Quick Picks}

Purchasers of Powerball and Mega Millions tickets have the option to select
the combination for each ticket they buy, and roughly 30\% of tickets are sold
with such self-selected combinations \cite{quick}.
But self-selection leads to a greater likelihood of collisions, where multiple
players pick the same combination and hence must share the prize pool should they win.
People tend to choose numbers that are meaningful to them, such as dates and arithmetic progressions.
This lack of independence skews certain combinations to be selected far more often
than would be suggested by chance \cite{uneven}.

The remaining 70\% of tickets for these lotteries are sold through {\em Quick Picks},
where the point-of-sale terminal generates a combination at random.
Details of the generation algorithm are not available to us, but we presume that
something like a standard linear congruential generator (LCG) is used to produce
pseudorandom numbers.
These generators iteratively produce a sequence of values using the recurrence relation
$$X_{n+1} = \left(a \cdot X_{n} + c\right) \mod m$$
Instantiated with appropriate constants $a$, $c$, and $m$,
one can permute through all the values of $m$ before repeating.
See Knuth \cite{Knuth-v2} for a thorough discussion on the theory of
random number generation.

Our presumption is that such methods do an effective job selecting tickets
with uniform probability on each sales terminal.
But under the well known {\em birthday paradox}, collisions occur surprisingly
early in any such independent sampling strategy.
We expect the first collision to happen after about $\sqrt{\frac{\pi N}{2}} \approx 1.25\sqrt{N}$ tickets sold, where $N$ is the size of the ticket space.
For Powerball, where $N = $ 292,201,338, this works out to an expected sales collision
after only 21,367 tickets are sold.

The problem of collisions is further complicated because tickets are
sold simultaneously at thousands of terminals across the nation.
Synchronization of the random number generators across these machines (with the
same constants $a$, $c$, $m$, and initial $X_0$) would be disastrous, because the
same combinations would get sold repeatedly by different stores.

Quick Pick works independently across different lottery terminals.
Kelly Cripe, spokeswoman for the
Multi-State Lottery Association which runs Powerball, stated that Quick Picks
``has no memory of what it previously selected'' as an explanation for why
multiple players can get identical combinations
\cite{rocheleau_2016}.
Presuming that the constants and initializations of the random number generators have been chosen correctly,
the collection of tickets across stores should be generated independently, with
the resultant collision probabilities well defined as a function of the number of tickets sold.

All of this leads to the question of whether it is possible to construct an efficient distributed lottery
scheme such that the probability of having to share the prize is minimized.

\section{Distributed Strategies to Generate Tickets}

We consider a setting where $m$ stores independently generate tickets on demand.
Each distinct lottery ticket can be ranked, or equivallently put in a bijection with a
distinct integer ranging from 0 to
$N-1$.
It is a straightforward task to unrank each such integer into a ticket, as well as
the inverse operation of ranking each ticket to corresponding integer using a
a recursive combinatoric approach. We discuss such operation in the appendix.


In our analysis, we consider a ticket as winning only if it claims part of the
grand prize, ignoring smaller prizes granted for similar but incomplete matches.
We assume the winning ticket will be drawn uniformly at random over the ticket space.
To simplify our analysis, we assume that that all tickets are bought through a
Quick Pick mechanism, meaning that customers cannot or do not
selected their own combinations.

Our goal here is to devise an efficient, distributed mechanism to implement Quick Pick
so as to optimize the expected value of a ticket, given that $n$ tickets have been sold.
We consider three different models:

\begin{itemize}

\item {\em Independent Generation} --
This is the simplest ticket generation strategy, and the one presumably implemented in current lottery point-of-sales terminals.
Each store generates a integer in the ticket space from $0$ to $N-1$ uniformly at
random on demand for each customer, which is unranked to generate an appropriate combination.
Equivalently, the process of selecting balls from an urn could be simulated to generate
tickets on demand.

Under such a system, each of the $m$ stores generate tickets independently, without memory of what they or any other store have generated in the past.
The downside is that no mechanism exists to prevent the same ticket being generated
twice, in different stores or even the same store.

\item
{\em Central Server Generation} --
At the other end of the spectrum, we consider a central server that stores communicate with, that ensures no duplicate ticket ever gets sold until the $(N+1)$st request.

Such a server could be implemented constructing a random permutation of the
entire ticket space, and respond to the $i$th ticket request with the $i$th element in this ordering.
Alternately, we can represent the ticket space as a bit vector, and search from a randomly selected position $0 \leq x \leq N-1$ to return the first open index $i \ge x$.
After $N$ tickets have been sold, every subsequent ticket sold after will result in a collision, but this is clearly unavoidable due to the pigeonhole principle.

Although this central server idea appears to be optimal in terms of preventing collisions, it requires constant communication between each sales terminal and the server.
If at any point the connectivity is lost, tickets cannot be dispensed.
We seek a generation approach where ticket machines can work independently,
without any need of external communication while still minimizing collisions.

\item
{\em Deterministic Pairing} --
Here we propose a strategy where each store is assigned a ``partner'', such that
each store and its partner comprise a pair.
Thus $m$ stores yield $p = \lceil m/2 \rceil$ pairs.
We partition the ticket space $N$ into $p$ regions, and assign a distinct region of
size $N/p$ to each pair.
This represents the range of tickets that a particular pair is allowed to sell from.
(Recall that each possible ticket is represented by a distinct integer from 0 to $N-1$.)

One of the stores in each pair sells tickets in increasing order from the front of the region, while the other store sells tickets in decreasing order from the back.
This guarantees that no collisions will occur for each pair until they exhaust
the entire region, making it optimal for two stores.
After the store partners and ticket ranges have been assigned, stores need not communicate further with
any external party. 

\end{itemize}

\section{Combinatorial Analysis}

We now determine the expected value of a purchased ticket,
given that $k$ tickets have been sold. Let $P$ be the prize pool for a winning ticket, and $N$ be the size of the ticket space. We first consider the expected value of a single ticket. If the contribution of this ticket is unique among all tickets sold, then its expected value is $P/N$, because the probability of it winning is $1/N$ and the prize if it wins is $P$. If the ticket is not unique and shares its numbers with $g$ tickets in total, then its expected value is $P/(gN)$ since the prize would now be split among $g$ people resulting in a prize of $P/g$.

To get the expected value of a purchased ticket, we sum over all the expected prizes for each ticket and divide by the total number of tickets. To find the sum of these expected prizes, we make the following observation. If we consider just a single set of $g$ tickets that share the same numbers, then the sum of the expected values for those tickets will always be $P/N$, regardless of $g$. This is due to the fact that for a given ticket number with $g$ collisions, each ticket in the same set will have expected value $P/(gN)$ and since we have $g$ of them, we get a total expected value sum of $P/(gN) * g = P/N$. Thus, to compute the expected value sum over all tickets, we simply need to count the number of distinct ticket numbers and multiply this value by $P/N$.

In summary, the expected value of each ticket is simply the number of distinct tickets sold multiplied by $P/(kN)$, where $k$ the number of tickets sold so far.



\subsection{Independent Random Generation}

In the case of independent random generation, the number of distinct tickets can be computed analogously to the number of distinct birthdays among a random sample of $k$ people.

This is a known problem, but to motivate the solution, we assume we know how to compute the number of distinct birthdays for $k-1$ people, and consider what happens when we add a new person $k$. The probability $k$ does not share a birthday with any of the original $k-1$ people is $\left(\frac{N-1}{N}\right)^{k-1}$, and we can simply increase the expected value by 1 in such a case. In the other scenario, the new person does not contribute to the number of distinct birthdays, so the value does not increase.

This is summarized by the following recurrence, where $N$ is the size of the ticket space:
\begin{align*}
E(k) &= \left(\frac{N-1}{N}\right)^{k-1}\left[1 + E(k-1)\right] + \left(1-\left(\frac{N-1}{N}\right)^{k-1}\right)\left[E(k-1)\right] \\
&= \left(\frac{N-1}{N}\right)^{k-1} + E(k-1)\\
&= \sum_{i=0}^{k-1} \left(\frac{N-1}{N}\right)^{i}   
= \frac{1 - \left(\frac{N-1}{N}\right)^{k}}{1 - \left(\frac{N-1}{N}\right)}\\
&= N \cdot \left(1 - \left(\frac{N-1}{N}\right)^{k}\right)
\end{align*}

To get the final expected value, we multiply by $P/(kN)$ to get:

$$E_{\IR}(k) = \left[1 - \left(\frac{N-1}{N}\right)^k\right] \cdot P / k$$

\subsection{Central Server}

In the case of the central server, each ticket contributes to a unique number. Thus, the expected number of distinct tickets is equal to the number of tickets sold until $N$ tickets are sold. At that point, the maximum number of distinct tickets are sold.
The expected value of each ticket is given by:

$$E_{\CS}(k) = 
\begin{cases} 
P/N & k \le N \\
\frac{P}{k} & k > N
\end{cases}
$$

\subsection{Deterministic Pairing}

The deterministic pairing scheme can be modeled by considering a balls and bins problem with a limited capacity $c$ for each bin, where the balls represent a ticket, and the bins represent the partitions of the ticket space.
A ball is discarded whenever it is thrown into a full bin. This captures the fact that the ticket values are recycled after a partition of the ticket space is all used up.
We calculate the total number of balls remaining in the bins after $k$ balls are thrown. This value of the number of balls remaining represents the number of distinct tickets sold so far.
In the worst case, all the balls get thrown into a single bin and $k-c$ balls are discarded, but this is highly unlikely. 

We solve this analytically for the case of two bins with $c = N/2$. This is equivalent to considering two pairs of stores with an even partition of the ticket space with size $N/2$ each. The expected value of random variable $X$ with outcome values $x_i$, each with probability $p_i$:
$$E(X) = \sum_{i=1}^{k} p_i x_i$$
In this setting, we wish to consider every possible sequence of ball tosses. We can represent this as a binary string, in which 0 represents the left bin and 1 represents the right bin. The $i$th bit of the string will represent the $i$th ball thrown. We consider all binary strings of length $k$, and compute the expected value by summing over the values when there are $i$ zeros and $k-i$ ones directly for all $i$. 

The values, in this case, are the minimum between $i$ (or $k-i$) and $N/2$. This reflects the discarding aspect as a bin cannot have more than $N/2$ balls. Thus, our final expected value, given $k$ tickets, is:

$$E_{\DP}(k) = \sum_{i=0}^{k} \frac{\binom{n}{i}}{2^k}\cdot \frac{N}{kP} \Big.\left[\min\left(i,\frac{N}{2}\right) + \min\left(k-i,\frac{N}{2}\right)\Big.\right]$$


\begin{figure}[!htb]
	\begin{center}
		\includegraphics[width=4.0in]{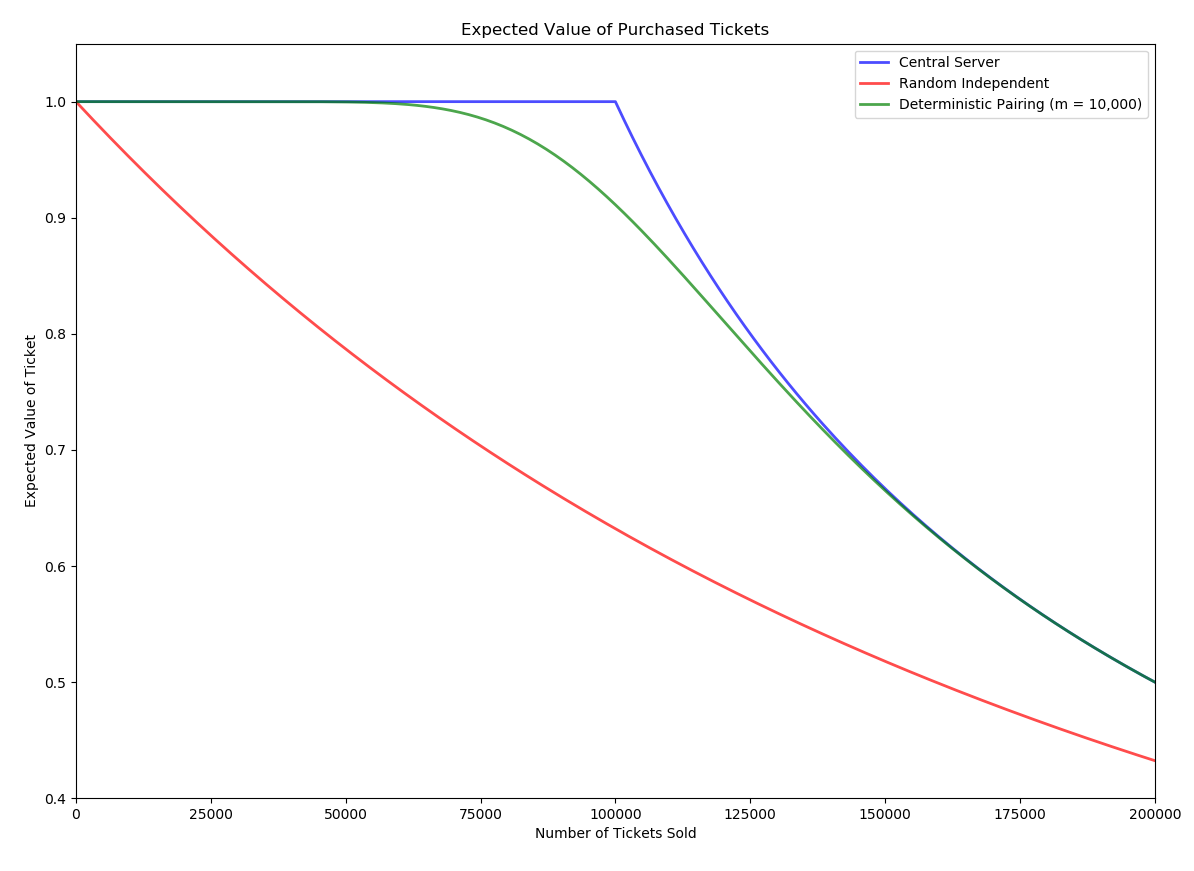}
	\end{center}
	\caption{The expected fraction of the pool size claimed over all tickets sold under
		three different models, for $N=100,000$.   The deterministic pairing model approaches
		the impractical central server model, while strictly dominating independent generation.}
	\label{final-expected-value}
\end{figure}

We provide simulation results in Figure \ref{final-expected-value}, for $N=100,000$.For the expected value of purchased tickets, we see that the central server method is the best (although impractical to implement) maximizes expected value because it guarantees that each ticket in the ticket space is sold at least once each before any combination repeats. The random independent strategy does the worst of the three methods as collisions arise relatively quickly. The deterministic pairing method does quite well as it nearly does as well as the ideal server model, making it best among practical methods.

\section{When to Buy to Maximize the Expected Value of a Ticket}

We now analyze at what point in the jackpot is the expected value of a ticket maximal. To do so, we first estimate the number of tickets sold for a given jackpot size. We do this by collecting data on lottery ticket sales across the United States for Powerball \footnote{\url{https://lottoreport.com/ticketcomparison.htm}}. By graphing the number of tickets sold as a function of the jackpot, we note that the curve is approximately quadratic. Thus, we run linear regression to find the best fit quadratic formula. If $T(j)$ is defined to be a function of the jackpot that outputs the number of tickets sold, we find that it is approximately:
$$T(j) = 278.36j^2 - 5364.95j + 10582740.74$$
Here $j$ is measured in terms of millions of dollars.

\begin{figure}[!htb]
	\begin{center}
		\includegraphics[width=4.0in]{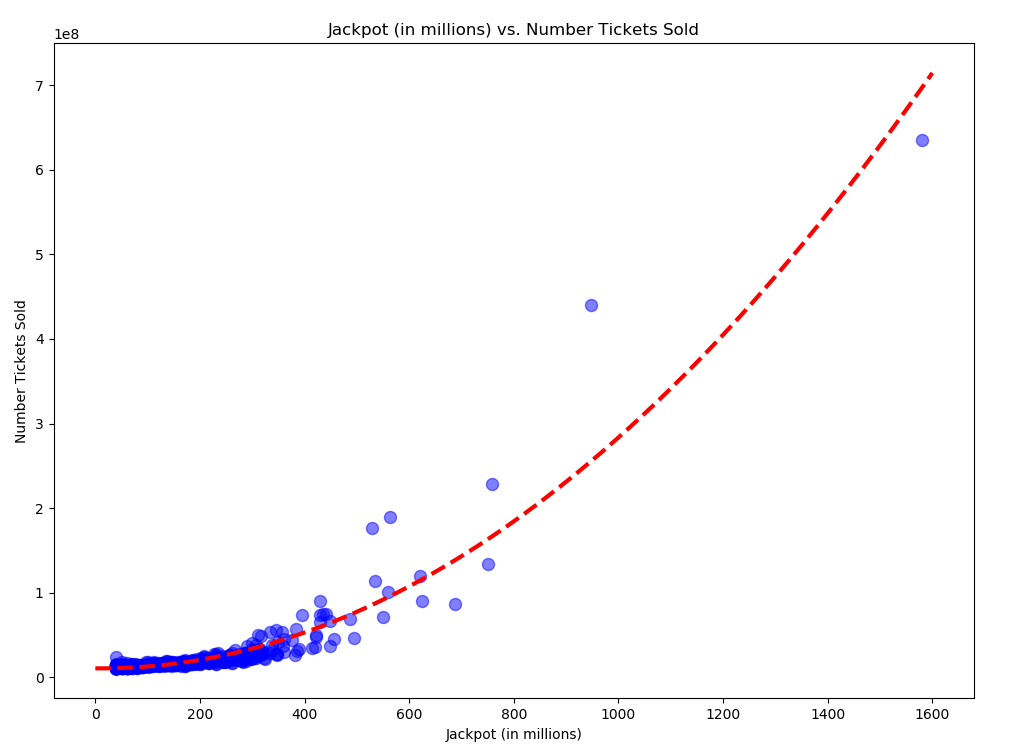}
	\end{center}
	\label{approx-num-tickets}
	\caption{The number of tickets sold as a function of the jackpot}
\end{figure}

Given this model, we can evaluate $E_{IR}(j)$, the expected value of a ticket for a given jackpot. Recall the expected value of a ticket can be computed for the random independent generation scheme, where $k = T(j)$ is the number of tickets sold and $N$ is the size of the ticket space (the size of the ticket space $N$ is 292,201,338 for Powerball), as

$$E_{IR}(j) = \left[1 - \left(\frac{N-1}{N}\right)^{T(j)}\right] \cdot P / T(j)$$

Similarly, for the (ideal) central server approach, we evaluate $E_{\CS}(j)$, as

$$E_{CS}(j) = 
\begin{cases} 
\frac{P}{N} & T(j) \le N \\
\frac{P}{T(j)} & T(j) > N
\end{cases}
$$

\begin{figure}
\begin{center}
\includegraphics[width=4.0in]{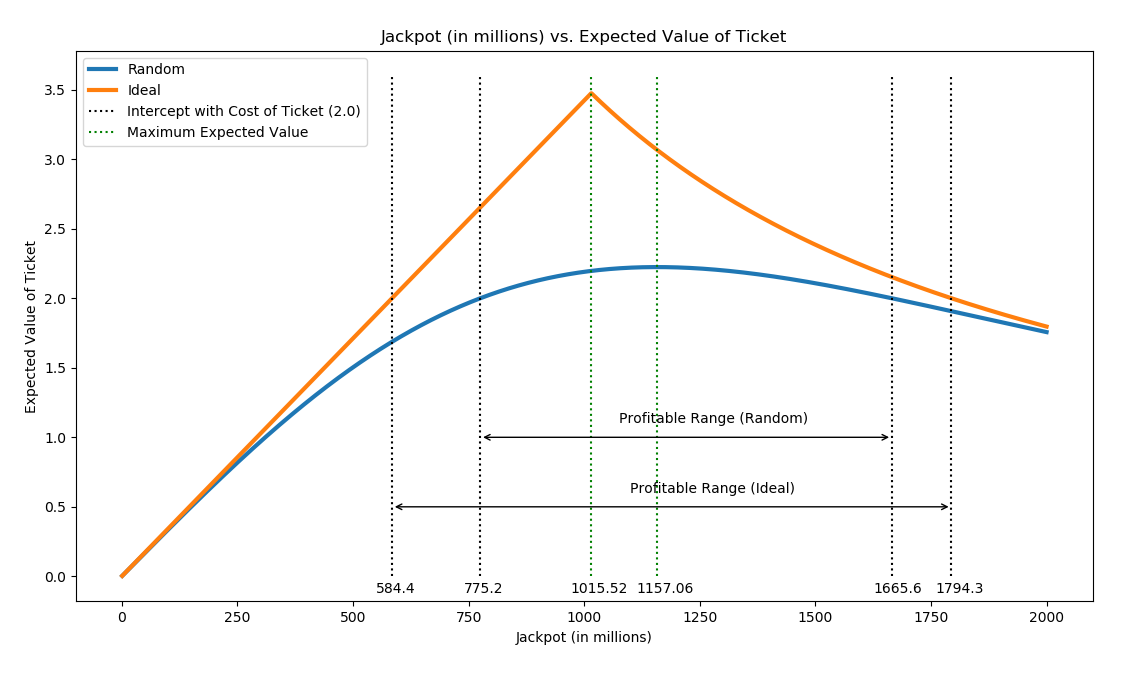}
\end{center}
\label{when-to-buy}
\caption{The expected value of a Powerball ticket under the Quick Pick/independent generation (blue) and maximally collision avoiding (red) sales models, assuming the number of tickets sold grows quadratically with the pool size.}
\end{figure}

In the end, we get the following results presented in Figure \ref{when-to-buy}. The cost of each lottery ticket is \$2, so we are interested in situations when the expected value of a ticket is greater than \$2. We see that for the standard Quick Pick scheme, one can expect to see returns when the jackpot is between \$775 million and \$1.67 billion. Here we see that a ticket provides its maximal return when the jackpot is around \$1.02 billion. But as the jackpot grows larger and larger, the expected number of tickets to be sold grows quadratically, and hence, the number of collisions overwhelm the returns of the jackpot. It becomes more and more likely that it will need to be shared among more people.

We note that under our proposed scheme, the range of the jackpot with positive expected returns is larger, between \$584 million and \$1.79 billion. As the pool size continues to increase, the expected value converges towards the standard Quick Pick method, but the expected value of our scheme is always larger than that of the standard Quick Pick method at all times, providing greater incentives for smart customers.

\section{Conclusion}

We propose an alternative to the standard ticket generation scheme used in popular lotteries that generally minimizes collisions and raises the expected value of a ticket. Our deterministic pairing method only requires an agreed setup between the lottery associate and its distributors. No further communication is required during sales. For future work, one may consider adding some degree of communication to establish how much more this method can be improved upon. Analyzing the impact of non-uniform ticket sales among stores (some more popular than others) is another factor to consider as well.

The reader may wonder what the catch is with our ticket generation procedure.
How can we really increase expected value by affecting sales strategy,
without any change in the cost of the lottery pool?
Over the course of a single lottery, it is clear that we accomplish our goals.
But there are certain subtleties in running a sequence of lotteries, where the pools increase
whenever there is no winner the previous week.
By reducing duplicates entries, we increase the likelihood that prize will be claimed each week.
Over a sequence of lotteries, our scheme will create fewer large pools resulting from long runs of unsuccessful contests.
But no one likes to share, and a lucky winner would be more likely to keep all of it.

\bibliography{myref}
\bibliographystyle{ieeetr}

\section{Appendix}

We describe a process of converting an integer \texttt{n} (rank) into a sequence of numbers (a lottery ticket). The reverse, going from ticket to rank, can be done in a similar manner, where the steps are nearly reversed. Recall that a lottery ticket consists of 5 integers from 1 to 69 (white balls), and a sixth integer ranging from 1 to 26 (the ``powerball''). In this case, \texttt{n} must be at least 0 and less than 292,201,338. We first consider generating just the white balls in the range from 0 to 68 (we simply add 1 to each ball in the end).

Our approach is to generate each number sequentially, keeping track of a lower bound to ensure a strictly increasing order. Let \texttt{GenTicket(n,l,s)} be a function where \texttt{n} is the rank in question, \texttt{l} is a lower bound for the numbers we are allowed to use, and outputs a sequence of \texttt{s} integers in strictly increasing order with values at least 0 and less than \texttt{h} (globally provided). To generate the white balls, we would call \texttt{GenTicket(n,0,5)}, where \texttt{n} is the rank and \texttt{h} is globally provided as 69. We can define \texttt{GenTicket} as follows, where \texttt{Binom(n,k)} counts the number of combinations to choose \texttt{k} out of \texttt{n} objects:

\begin{verbatim}
global h : upper bound of ticket numbers
function GenTicket(n,l,s):
  if s == 1:
    return [n + l]
  else:
    i = 1
    while n >= Binom(h - l - i, s - 1):
      n -= Binom(h - l - i, s - 1)
      i += 1
    return [l + i - 1] + GenTicket(n, l + i, s - 1)
\end{verbatim}

Intuitively, we use \texttt{Binom} to determine how many tickets there are starting with the given lower bound, and continuously reduce the ticket space until we know the range in which the first number should lie. Then, we can recursively compute the rest of the ticket.

For the remaining powerball, we simply divide the integer \texttt{n} by the total number of possible white balls (11,238,513 for Powerball) to get the powerball number. As long as \texttt{n} is within the possible number of tickets, this will compute the appropriate ``level'' in the ticket space.

 Note that the first ($n = 0$) ticket would be $(1,2,3,4,5,1)$ and the last ($n = 292,201,337$) ticket would be $(65,66,67,68,69,26)$. As an example, we show the computation to find the 100,000,000th ticket. We solve with 0-based indexing, so at the end, we increase each value by 1 to get the standard ticket values.
 
 We first divide $n = $ 100,000,000 by 11,238,513 to get the Powerball number, which is 8. Now, we find the remainder of $n$ when divided by 11,238,513 to get 10,091,896, which represents the rank that we have to compute the values of the white balls for. We go step by step through the process to show how intermediate values are determined.

\newpage
\begin{verbatim}
# Our starting point
(?, ?, ?, ?, ?, 8)
n: 10091896, lower bound: 0, size: 5

# We want to check if 0 can be our first number
# The number of tickets with 0 as the first number is Binom(68,4)
# But since 10091896 > Binom(68,4), we know it cannot be 0.
# We reduce to get a new n = 10091896 - Binom(68,4) = 9277511.

# We continue the process. We can rule out 1 as the first number
# because 9277511 > Binom(67,4): n = 9277511 - Binom(67,4) = 8511031
# ...repeat 24 iterations in total...
# Finally, we get that 75142 < Binom(44,4). Our ticket lies in this set.
# Since we iterated 24 times from 0, our first number must be 24.
# We recursively repeat the process.

(24, ?, ?, ?, ?, 8)
n: 75142, lower bound: 25, size: 4
# Since we have used 24, our new lower bound is 25.
# Our new size is 4 since we only have 4 more to fill now.
# 62801 = 75142 - Binom(43,3)
# ...total 7 iterations...
# 5436 < Binom(36,3)
# Thus, we add 7 more to the new lower bound.

(24, 32, ?, ?, ?, 8)
n: 5436, l: 33, size: 3
# 4841 = 5436 - Binom(35,2)
# ...total 13 iterations...
# 67 < Binom(22,2)

(24, 32, 46, ?, ?, 8)
n: 67, l: 47, size: 2
# 46 = 67 - Binom(21,1)
# 26 = 46 - Binom(20,1)
# 7 < Binom(19,1)

(24, 32, 46, 50, ?, 8)
n: 7, low: 51, size: 1
# Since size = 1, we can return low + n.

# Final result
(24, 32, 46, 50, 58, 8)

\end{verbatim}
In the end, we increment each value by 1 to have every value start at 1, and we find that the 100,000,000th ticket is \texttt{(25, 33, 47, 51, 59, 9)}.

\end{document}